\documentclass[aps,reprint,superscriptaddress,preprintnumbers,nofootinbib]{revtex4-2}   % using this instead makes \cref work for sections, but then it complains about references
\usepackage{graphicx}  % needed for figures
\usepackage{dcolumn}   % needed for some tables
\usepackage{bm}        % for math
\usepackage{amssymb}   % for math
\usepackage{amsmath}
\usepackage{slashed}
\usepackage{mathtools}
\usepackage[colorlinks]{hyperref}
\usepackage{cleveref}
\usepackage{xcolor}

\DeclarePairedDelimiter{\floor}{\lfloor}{\rfloor}

\allowdisplaybreaks

\newcommand{\cF}{\mathcal{F}}
\newcommand{\cM}{\mathcal{M}}
\newcommand{\cO}{\mathcal{O}}
\newcommand{\fa}{\mathfrak{a}}
\newcommand{\wa}{w\cdot\mathfrak{a}}

\begin{document}

% the following line is for submission, including submission to the arXiv!!
%\hspace{5.2in} \mbox{Fermilab-Pub-04/xxx-E}

\title{Classical gravitational scattering at $\mathcal{O}(G^2)$ and all orders in spin}
\title{Classical gravitational spinning-spinless scattering at $\mathcal{O}(G^2S^\infty)$}
%\input author_list.tex       % D0 authors (remove the first 3 lines
                             % of this file prior to submission, they
			     % contain a time stamp for the authorlist)
\author{Rafael Aoude}
\email[]{rafael.aoude@uclouvain.be}
\affiliation{Centre for Cosmology, Particle Physics and Phenomenology (CP3),\\
Universit\'{e} catholique de Louvain, 1348 Louvain-la-Neuve, Belgium}
\author{Kays Haddad}
\email[]{kays.haddad@physics.uu.se}
\affiliation{Department of Physics and Astronomy, Uppsala University, \\
Box 516, 75120 Uppsala, Sweden}
\affiliation{Nordita, Stockholm University and KTH Royal Institute of Technology, \\
Hannes Alfv\'{e}ns v\"{a}g 12, 10691 Stockholm, Sweden}
\author{Andreas Helset}
\email[]{ahelset@caltech.edu}
\affiliation{Walter Burke Institute for Theoretical Physics,
California Institute of Technology,\\ Pasadena, CA 91125, USA}
% (includes institutions and visitors)

\date{\today}

\begin{abstract}
	Making use of the recently-derived, all-spin, opposite-helicity Compton amplitude, we calculate the classical gravitational scattering amplitude for one spinning and one spinless object at $\mathcal{O}(G^2)$ and all orders in spin.
	By construction, this amplitude exhibits the spin structure that has been conjectured to describe Kerr black holes.
	This spin structure alone is not enough to fix all deformations of the Compton amplitude by contact terms, but when combined with considerations of the ultrarelativistic limit we can uniquely assign values to the parameters remaining in the even-in-spin sector.
	Once these parameters are determined, much of the spin dependence of the amplitude resums into hypergeometric functions.
	Finally, we derive the eikonal phase for aligned-spin scattering.
\end{abstract}


\preprint{ CP3-22-32 }
\preprint{ UUITP-24/22 }
\preprint{ CALT-TH-2022-018}

\maketitle

%%%%%%%%%%%%%%%%%%%%%%%%%%%%%%%%%%%%%%%%%%%%%%%%%%%%%%%
%%%%%%%%%%%% Begin Introduction %%%%%%%%%%%%%%%%%%%%%%%
%%%%%%%%%%%%%%%%%%%%%%%%%%%%%%%%%%%%%%%%%%%%%%%%%%%%%%%

\section{\label{sec:intro}Introduction}

Recent years have seen a large mobilization within the scattering amplitudes community towards describing the gravitational coalescence of compact objects.
This stems from the necessity for ever-more precise gravitational wave templates in current and upcoming gravitational wave observatories \cite{Punturo:2010zz,VIRGO:2014yos,LIGOScientific:2014pky,LISA:2017pwj,Reitze:2019iox,KAGRA:2020agh,Saleem:2021iwi}, and because scattering amplitudes are eminently suited to calculating classical observables in the post-Minkowskian (PM) expansion \cite{Damour:2017zjx,Antonelli:2019ytb,Kosower:2018adc,Maybee:2019jus,Bjerrum-Bohr:2019kec,Kalin:2019rwq,Kalin:2019inp,Cristofoli:2021vyo,Khalil:2022ylj}.
This huge effort has led to unprecedented precision in the PM description of spinless scattering \cite{Cheung:2018wkq,Bern:2019nnu,Bern:2019crd,Cheung:2020gyp,Kalin:2020mvi,Kalin:2020fhe,Herrmann:2021lqe,Herrmann:2021tct,DiVecchia:2021bdo,Brandhuber:2021eyq,Bern:2021dqo,Bern:2021yeh,Dlapa:2021npj,Dlapa:2021vgp}, tidal effects \cite{Cheung:2020sdj,Haddad:2020que,Kalin:2020lmz,Cheung:2020gbf,Bern:2020uwk,AccettulliHuber:2020dal,Aoude:2020ygw}, and radiation \cite{Herrmann:2021lqe,Herrmann:2021tct,DiVecchia:2021bdo,Cho:2021arx,DiVecchia:2022owy,DiVecchia:2022nna,Mougiakakos:2022sic}.

Yet another pertinent property affecting the motion of the constituents of a binary is their individual rotational angular momenta.
The connection between classical rotational angular momentum and quantum spin appearing in scattering amplitudes is by now well understood \cite{Vaidya:2014kza,Maybee:2019jus,Bern:2020buy,Aoude:2021oqj}.
Classical scattering at 1PM is known to all orders in the spin vectors for Kerr black holes \cite{Vines:2017hyw,Guevara:2018wpp,Guevara:2019fsj,Arkani-Hamed:2019ymq,Aoude:2020onz} and general spinning bodies \cite{Aoude:2021oqj}.
Dynamics at 2PM have been understood up to quartic order in spin \cite{Guevara:2017csg,Guevara:2018wpp,Chung:2018kqs,Damgaard:2019lfh,Bern:2020buy,Aoude:2020ygw,Liu:2021zxr,Kosmopoulos:2021zoq,Jakobsen:2021lvp,Jakobsen:2021zvh,Chen:2021qkk}.
Until recently, progress past quartic order at 2PM has been restricted owing partly to the absence of a physical opposite-helicity Compton amplitude above this spin order \cite{Arkani-Hamed:2017jhn}.
Several approaches have been taken to remedy these unphysicalities \cite{Chung:2018kqs,Falkowski:2020aso,Bautista:2021wfy,Chiodaroli:2021eug,Aoude:2022trd}.
Results including spin at 3PM have also begun to emerge \cite{Jakobsen:2022fcj,Alessio:2022kwv}.

Recently, refs.~\cite{Aoude:2022trd,Bern:2022kto} have pushed the state-of-the-art in the scattering of spinning objects at 2PM past the fourth order in spin.
In the former work, we applied the heavy on-shell variables of ref.~\cite{Aoude:2020onz} to focus on the classical limit of the Compton amplitude.
Doing so allowed us to determine the classical opposite-helicity Compton amplitude free of unphysical poles.
We subsequently evaluated the 2PM amplitude up to eighth order in spin, fixing nearly all contact terms by imposing the so-called {\it black hole spin structure assumption} on the 2PM amplitude; see \cref{eq:SpinStructureAssumption}.
Contrasting with our on-shell approach, the authors of ref.~\cite{Bern:2022kto} started from a local Lagrangian including operators at fifth order in spin.
This enabled them to bypass the issues with unphysical poles in the Compton amplitude and construct the 2PM Hamiltonian for two general spinning bodies up to the fifth order in spin.
The shift symmetry proposed in ref.~\cite{Bern:2022kto} to describe Kerr black holes is equivalent to imposing the black hole spin structure assumption, and the results of these two works are in agreement where there is overlap.

In this Letter, we present the amplitude at $\cO(G^{2})$, all orders in spin, and with an arbitrary mass ratio for spinning-spinless scattering based on the Compton amplitude determined in ref.~\cite{Aoude:2022trd}.
Many of the observations made about the 2PM amplitude at finite spin in ref.~\cite{Aoude:2022trd} are elucidated and shown to hold to all orders in spin.
Being in possession of this amplitude, examining its ultrarelativistic limit allows us to uniquely determine the remaining contact-term coefficients in the even-in-spin sector of the Compton amplitude.
The resulting 2PM amplitude is surprisingly simple, with much of the spin dependence resumming into hypergeometric functions.

Observables related to classical scattering involving a spinning object can be derived from amplitudes using a variety of methods \cite{Kosower:2018adc,Maybee:2019jus,Kalin:2019inp,Bern:2020buy}.
One approach passes through the eikonal phase \cite{Amati:1987uf,Guevara:2018wpp,Bern:2020buy,Bautista:2021wfy,Haddad:2021znf,Jakobsen:2021zvh,Adamo:2021rfq,Alessio:2022kwv}, which we present for aligned-spin scattering.

Let us begin by writing the opposite-helicity Compton amplitude in the form most convenient for our present analysis.

%%%%%%%%%%%%%%%%%%%%%%%%%%%%%%%%%%%%%%%%%%%%%%%%%%%%%%%
%%%%%%%%%%%% End Introduction %%%%%%%%%%%%%%%%%%%%%%%%%
%%%%%%%%%%%%%%%%%%%%%%%%%%%%%%%%%%%%%%%%%%%%%%%%%%%%%%%

%%%%%%%%%%%%%%%%%%%%%%%%%%%%%%%%%%%%%%%%%%%%%%%%%%%%%%%
%%%%%%%%%%%% Begin Setup %%%%%%%%%%%%%%%%%%%%%%%%%%%%%%
%%%%%%%%%%%%%%%%%%%%%%%%%%%%%%%%%%%%%%%%%%%%%%%%%%%%%%%

\section{\label{sec:Compton}Compton amplitude for heavy spinning particles}

The all-spin, opposite-helicity gravitational Compton amplitude at leading order in $\hbar$ was presented by the present authors in ref.~\cite{Aoude:2022trd}.
A vast majority of contact term contributions were fixed by imposing the black hole spin structure assumption on the 2PM amplitude, which demands that spin structures (parametrized in terms of the ring radii $\fa^{\mu}_{i}\equiv S^{\mu}_{i}/m_{i}$ for $S^{\mu}_{i}$ the classical spin vectors and $m_{i}$ the masses) appear in the combination
\begin{align}\label{eq:SpinStructureAssumption}
(q\cdot \mathfrak{a}_i)(q\cdot \mathfrak{a}_j) - q^2 (\mathfrak{a}_i \cdot \mathfrak{a}_j), \qquad i,j=1,2.
\end{align}
This assumption is equivalent to the shift symmetry imposed on the 2PM amplitude above $\cO(\fa^{3})$ in ref.~\cite{Bern:2022kto}.
In fact, requiring that the 2PM amplitude possesses this symmetry is equivalent to requiring that the opposite-helicity Compton amplitude is invariant under the shift\footnote{The inverse of the Mandelstam on the right-hand side is not strictly necessary. It allows $\xi$ to be dimensionless, but the shift can be expressed locally by omitting $s_{34}$ and making $\xi$ dimensionful. The same-helicity Compton amplitude is not invariant under this shift.}
\begin{align}\label{eq:ComptonShiftSymmetry}
    \fa^{\mu}\rightarrow\fa^{\mu}+\xi\frac{q_{3}^{\mu}+q_{4}^{\mu}}{s_{34}}.
\end{align}
At leading order in $\hbar$, the most general arbitrary-spin, opposite-helicity Compton amplitude invariant under this shift is (modulo the overall coupling)
\begin{align}
	\label{eq:CorrectedGR}
	\cM_{\text{cl}}^{s}&=e^{-\mathfrak{s}_{1}}\sum_{n=0}^{2s}\frac{1}{n!}\bar{K}_{n}+m^{2}(\wa)^4\mathcal{C},
\end{align}
where
\begin{align}
    \bar{K}_{n}&\equiv
    \begin{cases}
        K_{n}, & n\leq4, \\
        K_{4} L_{n-4}  - K_{3} \mathfrak{s}_{2} L_{n-5}, & n>4,
    \end{cases} \\
    \mathcal{C}&\equiv \sum_{n=0}^{2s-4}\sum_{j=0}^{\floor{(2s-4-n)/2}}d_{n,j}\mathfrak{s}_{1}^{n}(\mathfrak{s}^2_1 - \mathfrak{s}_{2})^{j} ,
    \label{eq:ContactTerms}
\end{align}
with
\begin{align}
    K_{n}&\equiv\frac{y^{4}}{s_{34}t_{13}t_{14}}\left(\frac{t_{14}-t_{13}}{y}w\cdot\mathfrak{a}\right)^{n},
    \\
    \quad L_{m} &\equiv \sum_{j=0}^{\floor{m/2}} \binom{m+1}{2j+1} \mathfrak{s}^{m-2j}_{1} (\mathfrak{s}^2_1 - \mathfrak{s}_{2})^{j},\quad 
\end{align}
and
\begin{align}
    \mathfrak{s}_1 &\equiv (q_3 - q_4 )\cdot \mathfrak{a} , \\
	\mathfrak{s}_{2} &\equiv -4 (q_3 \cdot \mathfrak{a})(q_4 \cdot \mathfrak{a}) + s_{34} \mathfrak{a}^{2}.
\end{align}
The spin structures $w\cdot\fa$, $\mathfrak{s}_{1}$, and $\mathfrak{s}_{2}$ are individually invariant under \cref{eq:ComptonShiftSymmetry}.
Therefore, \cref{eq:CorrectedGR} is itself manifestly invariant under this shift.

We have taken both graviton momenta to be outgoing and the initial massive momentum to be incoming.
The graviton labeled by $3$ carries negative helicity, and that labeled by $4$ carries positive helicity.
The momenta are grouped into the Mandelstam variables $s_{34}=(q_{3}+q_{4})^{2}$ and $t_{1i}=(p_{1}-q_{i})^{2}-m_{1}^{2}$.
Finally, we have defined the four-vector $w^{\mu}\equiv[4|\bar{\sigma}^{\mu}|3\rangle/2$ and $y\equiv2p_{1}\cdot w$.

We have rearranged the contact terms with unfixed coefficients compared to ref.~\cite{Aoude:2022trd}.
As a consequence, the unfixed coefficients here are different from those there.
The coefficients here contribute at order $\cO(\fa^{n+2j+4})$.
Furthermore, we have written $\bar{K}_{n\geq5}$ in terms of $K_{3}$ and $K_{4}$, as opposed to $K_{2}$ and $K_{3}$ as in ref.~\cite{Aoude:2022trd}, by using the recursion relation for $K_{n\geq4}$ presented there.
The infinite-spin amplitude is trivially found by taking $s\rightarrow\infty$ in \cref{eq:CorrectedGR,eq:ContactTerms}.

Having suitably reshuffled the all-spin, opposite-helicity Compton amplitude, we move now to evaluating the $\cO(G^{2})$ spinning-spinless amplitude to all orders in spin.

%%%%%%%%%%%%%%%%%%%%%%%%%%%%%%%%%%%%%%%%%%%%%%%%%%%%%%%
%%%%%%%%%%%% End Setup %%%%%%%%%%%%%%%%%%%%%%%%%%%%%%%%
%%%%%%%%%%%%%%%%%%%%%%%%%%%%%%%%%%%%%%%%%%%%%%%%%%%%%%%

%%%%%%%%%%%%%%%%%%%%%%%%%%%%%%%%%%%%%%%%%%%%%%%%%%%%%%%
%%%%%%%%%%%% Begin Amplitude %%%%%%%%%%%%%%%%%%%%%%%%%%
%%%%%%%%%%%%%%%%%%%%%%%%%%%%%%%%%%%%%%%%%%%%%%%%%%%%%%%

\section{\label{sec:allSpin}All-spin scattering}

The classically-relevant part of the one-loop $2\rightarrow2$ amplitude is encoded in the coefficients for triangle topologies, specifically those with one massive and two massless propagators in the loop \cite{Neill:2013wsa,Cheung:2018wkq,Bern:2019crd}.
We construct these coefficients out of the Compton amplitude in \cref{eq:CorrectedGR}\footnote{The same-helicity Compton amplitude does not contribute to classical scattering at 2PM when \cref{eq:SpinStructureAssumption} is exhibited \cite{Neill:2013wsa,Aoude:2022trd}.} and the three-point amplitude describing a Kerr black hole \cite{Arkani-Hamed:2017jhn,Guevara:2018wpp,Chung:2018kqs,Guevara:2019fsj,Arkani-Hamed:2019ymq,Aoude:2020onz} using generalized unitarity \cite{Bern:1994zx,Bern:1994cg,Bern:1997sc} (see also refs.~\cite{Bern:2020buy,Chen:2021qkk,Aoude:2022trd} for an outline of this method applied to the problem at hand).
In ref.~\cite{Aoude:2022trd} the present authors used this method to evaluate the 2PM amplitude for two spinning bodies to eighth order in spin, and to all orders in spin for a spinless probe in a Kerr background.
Considering only one object to be spinning, we present here the amplitude to all orders in spin for arbitrary mass ratios.

The scattering amplitude for a spinning particle with mass $m_1$ and ring radius $\mathfrak{a}_1$ and a spinless particle with mass $m_2$ has an even-in-spin and an odd-in-spin part:
\begin{align}\label{eq:AllSpinAmplitude}
	\cM_{2\text{PM}}= \frac{2G^2 \pi^2m_{1}^{2}m_{2}^{2}}{\sqrt{-q^2}} \left(  \cM_{2\text{PM}}^{\text{even}}+i \omega \mathcal{E}_{1} \cM_{2\text{PM}}^{\text{odd}} \right).
\end{align}
These different sectors are given by
\begin{widetext}
\begin{align}\label{eq:AmplitudeEven}
	&\mathcal{M}^{\rm even}_{2\text{PM}} =
	m_{1} \left[ 3 (5\omega^2 - 1) \cF_{0}
	+\frac{1}{4}(\omega^2 - 1)  \cF_{2} Q
	+ \frac{8\omega^4 - 8\omega^2 + 1}{\omega^2 - 1} \cF_{1} Q
	- \frac{1}{2} \cF_{2} V
	%\right.\right.\nonumber \\ & \left.\left.
	+ \sum_{k=1}^{\infty} \frac{(8\omega^4 - 8\omega^2 + 1) }{(\omega^2 - 1)^{k+1}} \frac{(-1)^{k} 2 \Gamma[k]}{\Gamma[2k]} \cF_{k-1} V^{k} 
	\right]	
	\nonumber \\ &
	- m_2 \left[
		- 3 ( 5 \omega^2 - 1) \sqrt{\pi}\cF_{-1/2}
		- \frac{3\sqrt{\pi}}{4}\cF_{1/2}Q 
		- \frac{1}{\omega^2 - 1} \cF_{1} Q + \frac{15\sqrt{\pi}}{4}\cF_{1/2}V
		\right. 
	\nonumber \\ & \left.
		+6\sqrt{\pi} \sum_{k=1}^{\infty} \frac{\omega^{2k}}{(\omega^2-1)^{k+1}}   \frac{(-1)^{k} \cF_{k-1} V^{k} }{\Gamma[2k+1]\Gamma[5/2-k]} \left[ \;_2F_1\left(\frac{1}{2}-k,-k;\frac{5}{2}-k;\frac{1}{\omega^{2}}\right)
		     %\right.\right. \nonumber \\ & \left.\left.
		- \left(k+\frac{3}{2}\right) \;_2F_1\left(\frac{3}{2}-k,-k;\frac{5}{2}-k;\frac{1}{\omega^{2}} \right) \right]
\right.\notag \\
&\left.-\frac{1}{64}(3Q^{2}+30QV+35V^{2})\sum_{k=0}^{\infty}c_{k}^{(0)}+\frac{1}{16}(Q+7V)(Q+V)\sum_{k=1}^{\infty}c_{k}^{(1)}-\frac{1}{64}(Q+V)^{2}\sum_{k=2}^{\infty}c_{k}^{(2)}
\right]  ,
\end{align}
for even spin powers and
\begin{align}\label{eq:AmplitudeOdd}
	\mathcal{M}^{\rm odd}_{2\text{PM}} &=
	- m_1 
\left[
	4 \cF_{1}+ \sum_{k=0}^{\infty} \frac{(2\omega^2 - 1)}{(\omega^2-1)^{k+1}} \frac{(-1)^{k} 8 \Gamma[k+1]}{\Gamma[2k+1]} \cF_{k} V^{k} 
	\right]
	\nonumber \\ &
	-m_2 \left.\left[
			\frac{15\sqrt{\pi}}{2}\cF_{1/2} 
			+ \sum_{k=0}^{\infty} \frac{\omega^{2k} }{(\omega^2 - 1)^{k+1}} \frac{4^{1-k} \cF_{k} V^{k}}{(1)_{k}(2k-1)}
	\right.\right. \nonumber \\ &
	\qquad\qquad\qquad\qquad\times
	\left.\left[ \;_2F_1\left(-\frac{1}{2}-k,-k;\frac{3}{2}-k;\frac{1}{\omega^{2}}\right)- \left(k + \frac{5}{2}\right)\;_2F_1\left(\frac{1}{2}-k,-k;\frac{3}{2}-k;\frac{1}{\omega^{2}} \right)
	\right]
\right],
\end{align}
\end{widetext}
for odd spin powers.
The transfer momentum is given by $q^{\mu}$.
We have defined $\omega\equiv v_{1}\cdot v_{2}$, $\mathcal{E}_{1}\equiv\epsilon^{\mu\nu\alpha\beta}v_{1\mu}v_{2\nu}q_{\alpha}\fa_{1\beta}$, $Q \equiv (q\cdot \mathfrak{a}_{1})^2 - q^2\, \mathfrak{a}_{1} ^{2}$, and $V \equiv q^2 (v_2\cdot \mathfrak{a}_{1})^2$.
The 2PM amplitude depends on $Q$ through the hypergeometric function\footnote{These hypergeometric functions are closely related to Bessel functions.} 
\begin{align}
	\cF_{j} \equiv \frac{1}{\Gamma[j+1]} \;_{0}F_{1}\left(j+1; \frac{Q}{4} \right) ,
\end{align}
while the unfixed contact term coefficients enter in
\begin{align}\label{eq:2PMContact}
	c^{(i)}_{k} &\equiv \frac{1}{4^{k}} (Q + V)^{k} \binom{2k}{k-i}\sum_{j=0}^{\infty} \Delta d_{2k,j} Q^{j}  ,
	\\
	\Delta d_{2k,j} &\equiv  d_{2k,j} + \frac{16(k-j)(2k+1)}{(2j+2k+4)!}.
\end{align}
Finally, the notation $(j)_{m}$ indicates the Pochhammer symbol.

In ref.~\cite{Aoude:2022trd} it was observed that the odd-in-spin parts of the spinning-spinless 2PM amplitude were uniquely fixed by imposing \cref{eq:SpinStructureAssumption}, up to $\cO(\fa^{7})$.
The results in this section demonstrate that the unfixed contact term coefficients in \cref{eq:ContactTerms} do not enter the odd-in-spin sector of the 2PM amplitude for spinning-spinless scattering at any order in spin.
It is actually easy to understand why this happens.
Parity-even contributions to the 2PM amplitude with odd powers of spin are re-expressible such that they contain exactly one Levi-Civita symbol.
However, the contact terms depend on three four-vectors, $q_{3}^{\mu}$, $q_{4}^{\mu}$ and $\fa_{1}^{\mu}$, which, after inserting the contact terms into the cut, become $q^{\mu}$, $p_{2}^{\mu}$, and $\fa_{1}^{\mu}$.
There are thus only three vectors that can be contracted into the Levi-Civita symbol, so any Levi-Civita symbol coming from contact terms in \cref{eq:ContactTerms} that has quenched Lorentz indices vanishes when one of the particles is not spinning.

Armed with this all-spin amplitude, we are in a position to make some statements about the unfixed even-in-spin coefficients in \cref{eq:ContactTerms}.

%%%%%%%%%%%%%%%%%%%%%%%%%%%%%%%%%%%%%%%%%%%%%%%%%%%%%%%
%%%%%%%%%%%% End Amplitude %%%%%%%%%%%%%%%%%%%%%%%%%%%%
%%%%%%%%%%%%%%%%%%%%%%%%%%%%%%%%%%%%%%%%%%%%%%%%%%%%%%%

%%%%%%%%%%%%%%%%%%%%%%%%%%%%%%%%%%%%%%%%%%%%%%%%%%%%%%%
%%%%%%%%%%%% Begin Fixing Contact Terms %%%%%%%%%%%%%%%
%%%%%%%%%%%%%%%%%%%%%%%%%%%%%%%%%%%%%%%%%%%%%%%%%%%%%%%

\section{\label{sec:contactTerms}Contact terms and high-spin resummation}

Ultimately, we would like to understand which set of contact terms in the Compton amplitude describes black hole physics.
Answering this question definitively requires performing a matching computation to a quantity that unambiguously describes black hole dynamics.
In the absence of such an object to which we can compare, we content ourselves with identifying sets of contact terms that impart special qualities to the 2PM amplitude.
\Cref{eq:SpinStructureAssumption} (equivalently the 2PM analog of \cref{eq:ComptonShiftSymmetry} \cite{Bern:2022kto}) is one such special property, already almost entirely eliminating contact-term freedoms.
The ultrarelativistic ($\omega\rightarrow\infty$) limit gives us another handle with which we can uniquely fix the parameters appearing in \cref{eq:AmplitudeEven}.

Expanded in powers of the spin, the full spinning-spinless 2PM amplitude can be written as\footnote{We caution the reader that the subscript $k$ here starts at $k=0$, while the analogous subscript in ref.~\cite{Aoude:2022trd} starts at 1.}
\begin{align}
    &\cM_{2\text{PM}}=G^{2}m^2_{1}m^2_{2}\frac{\pi^{2}}{\sqrt{-q^{2}}} \\
    &\times\sum_{n=0}^{\infty}\sum_{k=0}^{n}\left(M_{k}^{(2n
    )}+i\omega\mathcal{E}_{1}M_{k}^{(2n+1)}\right)Q^{n-k}V^{k},\notag
\end{align}
where the $M_{k}^{(i)}$ are form factors in the $\cO(\fa_{1}^{i})$ spin sector depending only on the masses and $\omega$.
We noted in ref.~\cite{Aoude:2022trd} that certain values of the remaining contact-term coefficients were suggested by the ultrarelativistic limit of the 2PM amplitude.\footnote{This limit was also considered in ref.~\cite{Bern:2022kto} to fix values for parameters not determined by the shift symmetry.}
It was observed there that up to $\cO(\fa^{6})$ it was possible to improve the ultrarelativistic limits of some even-in-spin form factors by choosing certain values for the remaining coefficients.\footnote{The ultrarelativistic scaling of the amplitude as a whole is not affected by the values of the coefficients.}
Inspecting \cref{eq:AmplitudeEven} we can see that this is always possible.

The contributions to \cref{eq:AmplitudeEven} from the contact terms always enter at $\cO(\omega^{0})$.
As such, no values for the coefficients in \cref{eq:ContactTerms} can affect the ultrarelativistic limits of the form factors with $k=0,1$.
However, in the absence of the $c_{k}^{(i)}$ in the last line of \cref{eq:AmplitudeEven}, the form factors with $k\geq2$ scale as $\cO(\omega^{-2})$ when $\omega\rightarrow\infty$, a behavior which is worsened by the $c_{k}^{(i)}$.
Requiring the best $\omega\rightarrow\infty$ behavior of all even-in-spin form factors is thus equivalent to setting
\begin{align}\label{eq:ContactCoefficientsEven}
	d_{2k,j}&=-\frac{16(k-j)(2k+1)}{(2j+2k+4)!},
\end{align}
which imposes $c_{k}^{(i)}=0$ (see \cref{eq:2PMContact}).
\Cref{eq:ContactCoefficientsEven} produces agreement between \cref{eq:CorrectedGR} and the classical limit of the opposite-helicity Compton amplitude of ref.~\cite{Arkani-Hamed:2017jhn} up to fourth order in spin; past this order, the latter possesses unphysical poles.

\Cref{eq:SpinStructureAssumption,eq:ContactCoefficientsEven} jointly endow the all-spin 2PM amplitude with a remarkably compact form.
The former condition allows for the spin dependence to be written only in terms of $Q$, $V$, and $\mathcal{E}_{1}$, relegating spin effects to the hyperplane orthogonal to $q^{\mu}$, up to subleading-in-$\hbar$ effects.
Moreover, apart from the $\cO(V^{0})$ portion of the even-in-spin sector, the latter condition causes the resummation of all $Q$ dependence into hypergeometric functions.

Further still, at $\cO(V^{k})$ for fixed $k\geq 2$ ($1$) in the even(odd)-in-spin sector, all $Q$ dependence is encapsulated in precisely one $\cF_{j}$.
This explains an observation made in ref.~\cite{Aoude:2022trd} in the odd-in-spin sector that certain form factors at different spin orders are proportional to each other.
It also shows that this proportionality exists in the even-in-spin sector as well, specifically for the form factors whose ultrarelativistic behavior is improved by \cref{eq:ContactCoefficientsEven}.
The constants of proportionality can be obtained by expanding the amplitude in $Q$ at a fixed order in $V$:
\begin{subequations}\label{eq:HSUniversality}
\begin{align}
    M_{k}^{(2n+1)}&=\frac{4^{k-n}M_{k}^{(2k+1)}}{(1)_{n-k}(k+1)_{n-k}}, \quad 1\leq k\leq n,\label{eq:HSUniversalityOdd} \\
    M_{k}^{(2n)}&= \frac{4^{k-n}M_{k}^{(2k)}}{(1)_{n-k}(k)_{n-k}},\qquad\quad 2\leq k\leq n.\label{eq:HSUniversalityEven}
\end{align}
\end{subequations}
Because of the lower bounds on $k$, these proportionalities can only be observed when spin orders higher than four are considered.
We correspondingly dub the resulting resummation the \textit{high-spin resummation}.

Fixing the coefficients in \cref{eq:ContactTerms} in the odd-in-spin sector requires consideration of $\cO(\fa_{1}^{2n+1}\fa_{2}^{i>0})$ sectors of the 2PM amplitude.
The possibility of non-vanishing odd-in-spin coefficients is itself interesting: it implies that certain properties of the object described by those values can only be probed at $\cO(G^{2})$ in the classical limit by scattering with another spinning body.
Such a phenomenon is actually not novel, as we observed in ref.~\cite{Aoude:2022trd} that some coefficients imposing \cref{eq:SpinStructureAssumption} are left unfixed by the spinning-spinless sector of the scattering.
Similar behavior can also be seen in the spinning tidal results of ref.~\cite{Aoude:2021oqj}.

%%%%%%%%%%%%%%%%%%%%%%%%%%%%%%%%%%%%%%%%%%%%%%%%%%%%%%%
%%%%%%%%%%%% End Fixing Contact Terms %%%%%%%%%%%%%%%%%
%%%%%%%%%%%%%%%%%%%%%%%%%%%%%%%%%%%%%%%%%%%%%%%%%%%%%%%

%%%%%%%%%%%%%%%%%%%%%%%%%%%%%%%%%%%%%%%%%%%%%%%%%%%%%%%
%%%%%%%%%%%% Begin Eikonal Phase %%%%%%%%%%%%%%%
%%%%%%%%%%%%%%%%%%%%%%%%%%%%%%%%%%%%%%%%%%%%%%%%%%%%%%%

\section{Aligned-spin eikonal phase}

The eikonal phase allows for the relation of the amplitude to observables such as the linear impulse, the spin kick, and, when the motion is planar, the scattering angle \cite{Amati:1987uf,Bern:2020buy,Jakobsen:2021zvh}.
When the scattering of spinning objects is under consideration, the condition of planar motion is satisfied when both spin vectors are orthogonal to the plane formed by the impact parameter $\boldsymbol{b}$ and the asymptotic center-of-mass three-momentum $\boldsymbol{p}$.
For brevity, we will present the eikonal phase in this setup.

In terms of the amplitude, the eikonal phase at $\cO(G^{2})$ is
\begin{align}
    \chi&=\frac{1}{4m_{1}m_{2}\sqrt{\omega^{2}-1}}\int\frac{d^{2}\boldsymbol{q}}{(2\pi)^{2}}e^{i\boldsymbol{b}\cdot\boldsymbol{q}}\cM_{\text{2PM}}.
\end{align}
For aligned-spin scattering, the amplitude in \cref{eq:AllSpinAmplitude} greatly simplifies because $V=0$.
Further imposing \cref{eq:ContactCoefficientsEven}, the aligned-spin eikonal phase is a sum of two terms:
\begin{widetext}
\begin{align}
    \chi_{\text{as}}&=-\frac{G^{2}\pi m_{1}m_{2}}{4bx(1-x^{2})^{3/2}(\omega^{2}-1)}\left(\frac{\chi_{\text{as}}^{\text{even}}}{x\sqrt{(1-x^{2})(\omega^{2}-1)}}+\omega \chi_{\text{as}}^{\text{odd}}\right),
\end{align}
where
\begin{align}
    \chi^{\text{even}}_{\text{as}}&=-2 m_1 \sqrt{1-x^2} \left[x^2 \omega ^2 \left(x^2 \omega ^2+6 \left(\omega
   ^2-1\right)\right)+\left(1-\left(1-x^2\right)^{3/2}\right) \left(\omega
   ^2-1\right)^2\right]\notag \\
   &\qquad+m_2 x^2 \left[3 \left(\omega ^2-1\right) \left(5 \left(x^2-1\right) \omega
   ^2-2 x^2+1\right)-2 x^2 \sqrt{1-x^2}\right], \\
    \chi_{\text{as}}^{\text{odd}}&=8m_{1}\left[\omega^{2}x^{2}+(\omega^{2}-1)\left(1-(1-x^{2})^{3/2}\right)\right]+3x^{2}m_{2}\left[2+5(\omega^{2}-1)\sqrt{1-x^{2}}\right].
\end{align}
\end{widetext}
We've defined $x\equiv a_{1}/b$.
The unstylized $a_{1}$ and $b$ represent the magnitudes of the spatial spin and impact-parameter vectors, respectively.
The sign on the eikonal phase in the odd-in-spin sector depends on the direction of the spin vector.
The spinless-probe limit of this eikonal phase agrees with the first line of eq.~(11) in ref.~\cite{Siemonsen:2019dsu} for all non-negative powers of the spin.

%%%%%%%%%%%%%%%%%%%%%%%%%%%%%%%%%%%%%%%%%%%%%%%%%%%%%%%
%%%%%%%%%%%% End Eikonal Phase %%%%%%%%%%%%%%%
%%%%%%%%%%%%%%%%%%%%%%%%%%%%%%%%%%%%%%%%%%%%%%%%%%%%%%%

%%%%%%%%%%%%%%%%%%%%%%%%%%%%%%%%%%%%%%%%%%%%%%%%%%%%%%%
%%%%%%%%%%%% Begin Conclusion %%%%%%%%%%%%%%%
%%%%%%%%%%%%%%%%%%%%%%%%%%%%%%%%%%%%%%%%%%%%%%%%%%%%%%%

\section{\label{sec:conclusion}Summary}

Exploiting the cured opposite-helicity Compton amplitude derived in ref.~\cite{Aoude:2022trd}, we have presented in \cref{eq:AllSpinAmplitude,eq:AmplitudeEven,eq:AmplitudeOdd} the 2PM amplitude describing the scattering of a spinning and a spinless body.
By construction, the amplitude exhibits \cref{eq:SpinStructureAssumption}---a structure observed in Kerr black hole scattering at low spin orders \cite{Holstein:2008sx,Guevara:2017csg,Damgaard:2019lfh,Bern:2020buy,Kosmopoulos:2021zoq,Chen:2021qkk}---to all spin orders.
Our result includes the most general set of contact terms which adheres to \cref{eq:SpinStructureAssumption}, and demonstrates explicitly that these contact terms do not contribute at any odd spin order in spinning-spinless scattering at 2PM.

Analyzing \cref{eq:AmplitudeEven}, we noticed that it is always possible to improve the ultrarelativistic behavior of even-in-spin form factors by selecting appropriate values for the coefficients in \cref{eq:ContactTerms}.
These values are given in \cref{eq:ContactCoefficientsEven}.
Additionally, these values for the coefficients lead to a compact resummation of nearly all $Q$ dependence of the amplitude into hypergeometric functions.
At finite spin, this resummation is signalled by the proportionality of form factors at different spin orders---see \cref{eq:HSUniversality}.

Finally, we presented the eikonal phase in the aligned-spin scattering setup to all orders in spin and for a general mass ratio.
This quantity can be easily converted to observables pertinent to the process \cite{Amati:1987uf,Bern:2020buy,Jakobsen:2021zvh}.

We have found a compact form for the $\cO(G^{2}S^{\infty})$ spinning-spinless amplitude by imposing only two constrains upon it.
Despite the elegance of the result, its relevance to Kerr black hole scattering remains to be elucidated.
The ubiquity of hypergeometric functions in \cref{eq:AmplitudeEven,eq:AmplitudeOdd} is intriguing; ref.~\cite{Castro:2010fd} connected the hypergeometric functions solving the radial part of the Teukolsky equation to a hidden conformal symmetry in the near region of a Kerr black hole.
Investigation of this connection may provide hints as to the Compton/2PM amplitude that truly describes a Kerr black hole.

The remaining coefficients in \cref{eq:ContactTerms} that are not determined by \cref{eq:ContactCoefficientsEven} require an analysis of the spinning-spinning amplitude.
We leave this for future work.

%%%%%%%%%%%%%%%%%%%%%%%%%%%%%%%%%%%%%%%%%%%%%%%%%%%%%%%
%%%%%%%%%%%% End Conclusion %%%%%%%%%%%%%%%%%%%%%%%%%%%
%%%%%%%%%%%%%%%%%%%%%%%%%%%%%%%%%%%%%%%%%%%%%%%%%%%%%%%

\acknowledgements
Four-vector manipulations were performed using \texttt{FeynCalc} \cite{MERTIG1991345,Shtabovenko:2016sxi,Shtabovenko:2020gxv}.
We thank Clifford Cheung for insightful discussions.
We are grateful to Justin Vines for comments on this manuscript.
RA's research is funded by the F.R.S-FNRS project no. 40005600. RA wishes to thank the Mani L. Bhaumik Institute for their hospitality.
KH is supported by the Knut and Alice Wallenberg Foundation under grant KAW 2018.0116 ({\it From Scattering Amplitudes to Gravitational Waves}) and the Ragnar S\"{o}derberg Foundation (Swedish Foundations’ Starting Grant).
KH is grateful to Nordita for their hospitality.
AH is supported by the DOE under grant no. DE- SC0011632 and by the Walter Burke Institute for Theoretical Physics.

\bibliography{bibliographyAllSpin}

\end{document}